\documentclass[aps,prb,preprint]{revtex4}
\usepackage{graphicx}

\begin{document}
\title{Highly efficient interfacing of guided plasmons and photons in nanowires}
\author{Xue-Wen Chen, Vahid Sandoghdar, Mario Agio}
\affiliation{Laboratory of Physical Chemistry
and OptETH, ETH Zurich, 8093 Zurich, Switzerland}

\begin{abstract}
Successful exploitations of strongly confined surface plasmon-polaritons
critically rely on their efficient and rapid conversion to lossless channels.
We demonstrate a simple, robust, and broad-band butt-coupling technique
for connecting a metallic nanowire and a dielectric nanofiber.
Conversion efficiencies above 95\% in the visible and
close to 100\% in the near infrared can be achieved with realistic
parameters. Moreover, by combining butt-coupling with nanofocusing,
we propose a broad-band high-throughput near-field optical
microscope.
\end{abstract}

\maketitle

Confined surface plasmon-polaritons (SPPs) have enormous potential for manipulating
electromagnetic fields at optical frequencies and nanoscopic length
scales\cite{barnes03}, with applications for
interconnects\cite{conway07}, field-enhanced spectroscopy and
microscopy\cite{keilmann99,babadjanyan00,stockman04,yeo09},
sensing\cite{sharma07}, and quantum optics\cite{chang06,chang07}.
However, absorption of energy at optical frequencies by real metals makes
propagation of confined SPPs very lossy in comparison with the transmission
of photons in dielectric guides\cite{conway07,novotny94}.
Therefore, practical device
proposals require a rapid and efficient conversion of SPPs into
photons. In case of SPPs at planar interfaces, evanescent wave and
grating couplers are very effective\cite{raether88}. Thus,
researchers have extended these schemes to guided modes of
dielectric fibers (DF) and metallic wires (MW) in a side by side
arrangement\cite{chang06,lee08}, via
tapers\cite{keilmann99,bouhelier03,janunts05,ding07}, or by etching
a grating on a MW nanocone\cite{ropers07}. However, efficient
implementation of these approaches requires an interaction length
greater than a wavelength\cite{snyder83}, making losses an important
issue. Furthermore, the critical dependence of evanescent coupling
on the overlaps between the DF and MW modes, and the inherent
wavelength dependence of grating coupling limit their bandwidths. In
this Letter, we investigate the most straightforward and practical
way of interfacing guided SPPs and photons between a MW and a nano
DF\cite{tong03} in an axially-symmetric butt-coupling scheme.
We investigate the conversion process of photons to SPPs as a
function of wavelength and material, identifying the
moulding of SPPs at the coupling interface as the condition for reaching
efficiencies above 95\% in the visible and close to 100\% in
the near-infrared range. Furthermore, we present a practically feasible scheme to
overcome the long-standing conflict between strong field enhancement
and high throughput in scanning near-field optical
microscopy\cite{novotny95b}, and to couple a quantum emitter to a
propagating optical mode\cite{domokos02}.

\begin{figure}
\includegraphics[width=13cm]{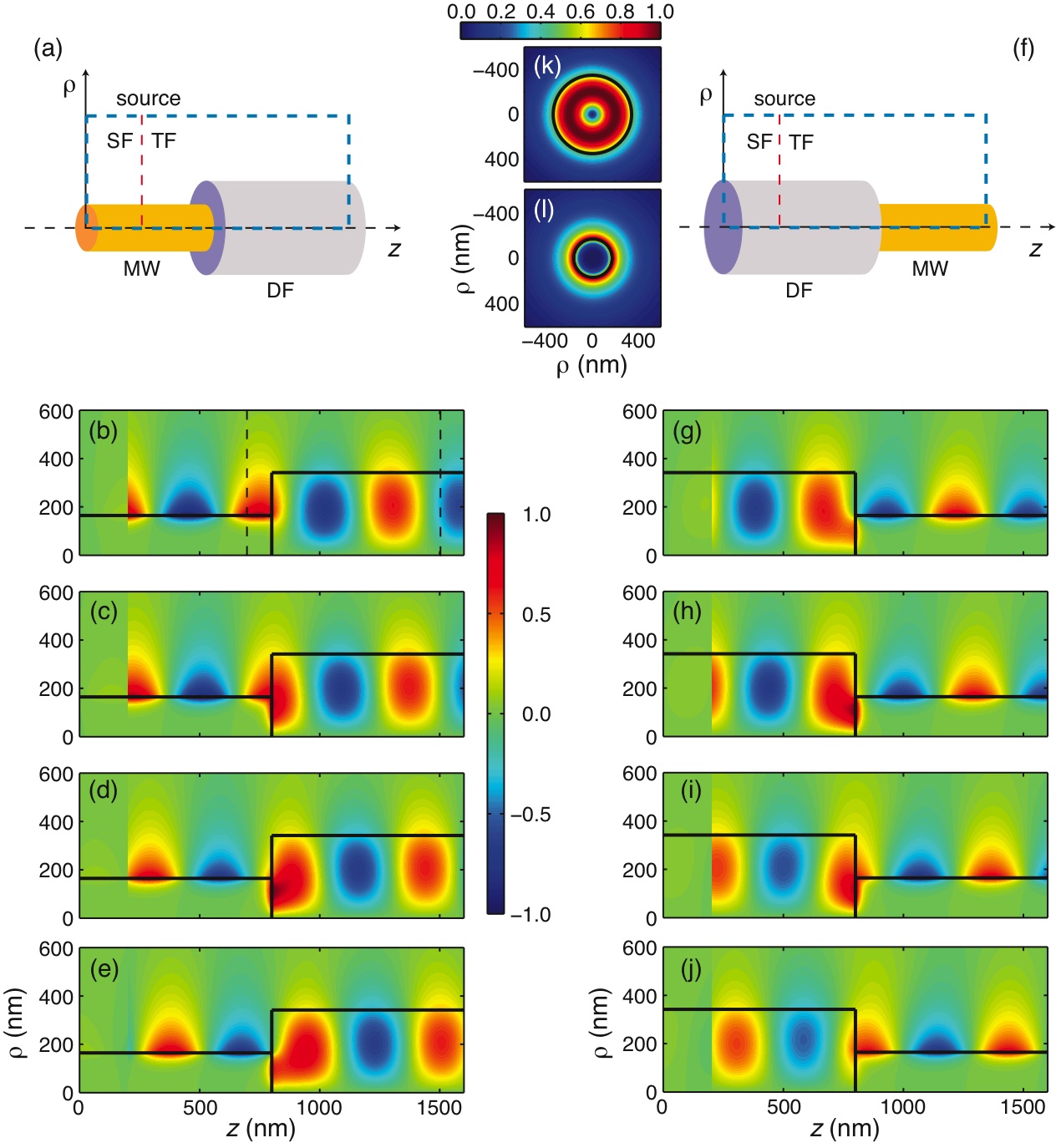}
\caption{(a) and (f) Layout of the coupling geometry. The vertical dashed
red line indicates the separation between the total-field (TF) and
the scattered-field (SF) region, and the position of the source. The
area delimited by dashed lines is used in the FDTD simulations.
(b)-(e) Snapshots of the
magnetic field at times $t_0$, $t_0+\tau$, $t_0+2\tau$, and
$t_0+3\tau$ respectively, for $\tau= 0.25$~fs. The vertical dashed
lines in (b) mark the area shown in \ref{fig2}(a). (g)-(j),
Snapshots for coupling from the opposite direction. (k) and (l),
Magnetic field patterns of the modes in the DF and MW,
respectively. The radii of the silica DF and silver MW are 342~nm
and 164~nm, respectively, and the wavelength was set at
$\lambda$=633~nm.\label{fig1}}
\end{figure}

\begin{figure}
\includegraphics[width=13cm]{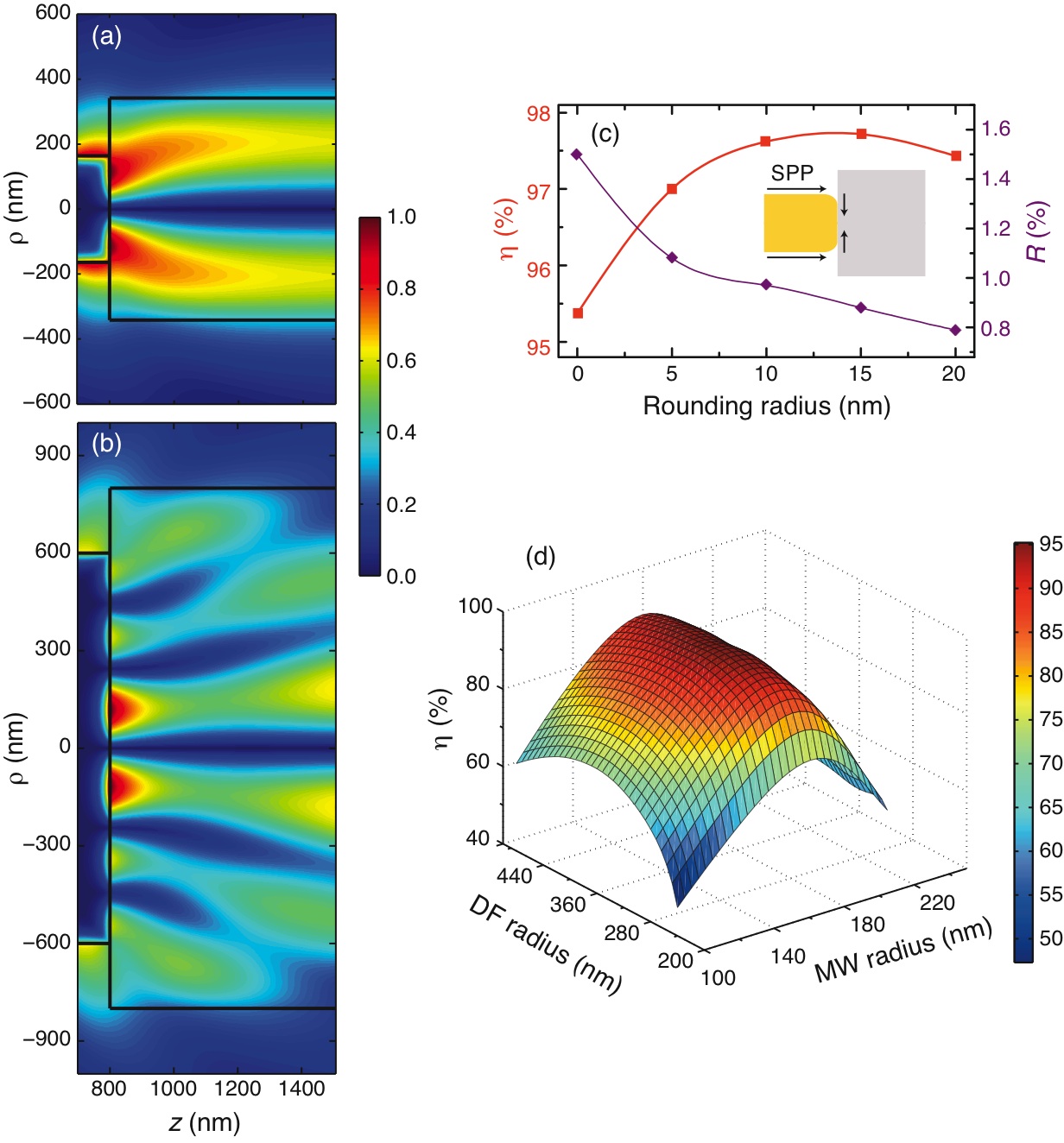}
\caption{(a) and (b) Time-averaged magnetic
field of a silver MW coupled to a silica DF for two different sets of
radii. See text for details. (c) The conversion efficiency $\eta$
and reflectivity $R$ as a function of the rounding radius of the
silver MW for the parameters of \ref{fig1}.
The inset illustrates the geometry and the
bending of SPPs onto the interface. (d) $\eta$ as a function
of the MW and DF radii. The wavelength was set to
$\lambda$=633~nm.\label{fig2}}
\end{figure}

We performed body-of-revolution finite-difference time-domain (FDTD)
calculations\cite{taflove05} to trace the propagation of the optical
signal along the guides in a rigorous and computationally efficient
manner, and to acquire an intuitive understanding of the mode conversion.
\ref{fig1}(a) and (f) sketch the geometrical arrangements for
interfacing SPPs on a silver\cite{CRChandbook} MW to guided photons
in a silica DF and vice versa for a vacuum surrounding.
The FDTD computational domain is restricted to the area enclosed
by the dashed lines in \ref{fig1}(a) and (f), and its radial dimension
is never less than two wavelengths.
In each case, a steady-state SPP or a TM$_{01}$ mode is launched
from the left-hand side using the total-field/scatter-field
technique\cite{taflove05}. 
The dashed red lines in \ref{fig1}(a) and (f) mark the source,
which separates the scattered-field from the total-field region.
The Drude model with conductivity is used to account for the
dispersive properties of metals. The parameters are deduced by
fitting the tabulated experimental data\cite{CRChandbook,palik98}
across a small spectral range (100~nm).
\ref{fig1}(k) and (l) depict the
time-averaged transverse magnetic field profiles of the
radially-polarized modes at the vacuum wavelength of
$\lambda=633$~nm in the DF and MW with radii of 342~nm and
164~nm, respectively. The snapshots of
the magnetic field in \ref{fig1}(b)-(e) illustrate how SPPs travel
on the surface of the MW and are converted to photons in the
TM$_{01}$ mode of the DF . To determine the conversion
efficiency $\eta$, we computed the ratio between the transmitted
power in the guided mode right after the MW-DF interface and the
incident power at the position right before it. The latter was
calculated separately for an infinitely long MW or DF using the same
source. In the example of \ref{fig1}(a), the MW and DF
yielded $\eta$=95\%.
We used a 2~nm grid here and found that the relative error between 1~nm
and 2~nm grids is within 0.5\%.
A grid size of 0.5~nm is used for \ref{fig2}(c) and \ref{fig5}.
\ref{fig1}(g)-(j) show the reverse situation where light
originates in the silica DF and is converted to SPPs of the silver
MW with the same efficiency as in the previous case. The great
advantage of the butt-coupling scheme is that the conversion between
SPPs and guided photons takes place across an interface. This
feature minimizes the impact of propagation losses,
which significantly increase as wavelength
and MW radius decrease\cite{novotny94}. For instance,
the MW at the wavelength considered in \ref{fig1}
would yeald losses as high as 0.72~dB/$\mu$m.

To explore the underlying coupling mechanism, in \ref{fig2}(a) we
plot the time-averaged magnetic field for the
region between the dashed lines in \ref{fig1}(b). We find that
the SPP moulds around the end of the MW (see arrows in the inset of
\ref{fig2}(c)) in such a way that the magnetic field assumes a
maximum value at a certain distance from the axis on the MW-DF
interface. Here, we tuned the radii of the two guides to obtain an
SPP field profile that optimizes its coupling to the TM$_{01}$ mode of
the DF. \ref{fig2}(b) displays another example of the field
distribution for larger MW and DF radii of 600~nm and 800~nm,
respectively. In this case the SPP magnetic field has three maxima,
which clearly shows that the SPPs interfere and form a
standing wave at the interface.
Thus, a simple analysis based on mode-matching of the two
guides\cite{snyder83} of \ref{fig1}(a) only predicts $\eta$=87\%
because it fails to account for the behaviour of SPPs.
In fact, \ref{fig2}(c) shows that
rounding off the MW edges facilitates this SPP folding process,
improving $\eta$ and reducing reflections. Having demonstrated the
possibility of very high conversion efficiency between SPPs and
photons, it is important to examine the feasibility and reliability
of this scheme for laboratory and technological applications. An
important issue of concern is the fabrication tolerance.
\ref{fig2}(d) displays $\eta$ as a function of the MW and DF
radii at $\lambda=633$~nm, revealing that conversion efficiencies
greater than 90\% can be achieved even if variations of up to 15\%
take place for the radii. Moreover, we have verified that even a
structure with an air gap of up to 50~nm between the MW and DF would
yield a conversion efficiency greater than 94\%.

\begin{figure}
\includegraphics[width=6cm]{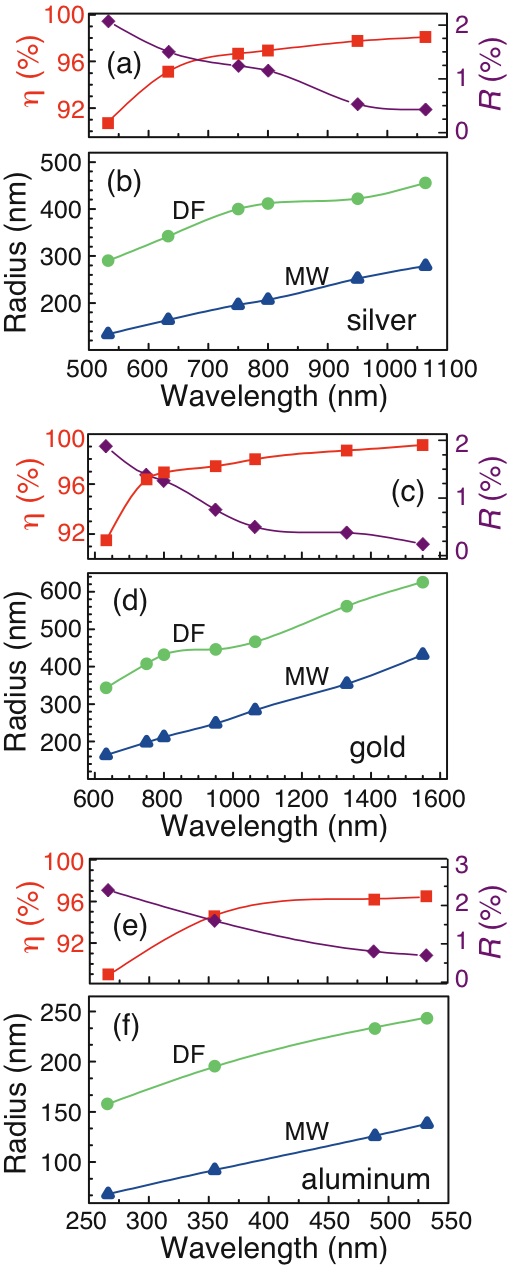}
\caption{Coupling efficiency and optimal radii of the MW and DF
as a function of wavelength.
(a) and (b) silver MW, (c) and (d) gold MW, (e) and (f) aluminium MW.
\label{fig3}}
\end{figure}

Next, we investigated the application of silver, gold\cite{CRChandbook} or
aluminium\cite{palik98} MW for operation at various wavelengths of
interest in the visible,
near infrared, and ultraviolet spectral regimes. We first
found the radii by maximizing $\eta$ for a given wavelength
using a mode-matching approach\cite{snyder83} and then applied the
FDTD technique to optimize them in a rigorous manner, by scanning
the radii around the previously estimated values.
\ref{fig3}(a) shows that for the case of silver MW, $\eta$
increases from 91\% to 98\%, while reflection $R$ decreases from
2.1\% to 0.4\% as the wavelength grows over a spectral range greater
than 500~nm. \ref{fig3}(b) plots the values of the silver-MW
and silica-DF radii corresponding to each optimized case. The results for
the opposite propagation direction are very similar (not shown).
\ref{fig3}(c) and (d) show that if silver is replaced by gold,
$\eta$, $R$ and the corresponding radii yield similar values and trends.
We note that around the telecommunication
wavelength of $\lambda=$1550~nm, a gold MW transmits nearly 100\% of
the power to the DF. \ref{fig3}(e) and (f) display the same
analysis for an aluminium MW with emphasis on the ultraviolet
region, where a rapid conversion of SPPs and photons becomes more
critical due to the very short SPP propagation length. We find that
even at a wavelength as short as $\lambda=266$~nm, $\eta$ can reach
89\% if the DF and MW radii are respectively chosen to be 158~nm
and 68~nm, for which the SPP propagation loss in the MW amounts
to 3.62~dB/$\mu m$.

\begin{figure}
\includegraphics[width=5.5cm]{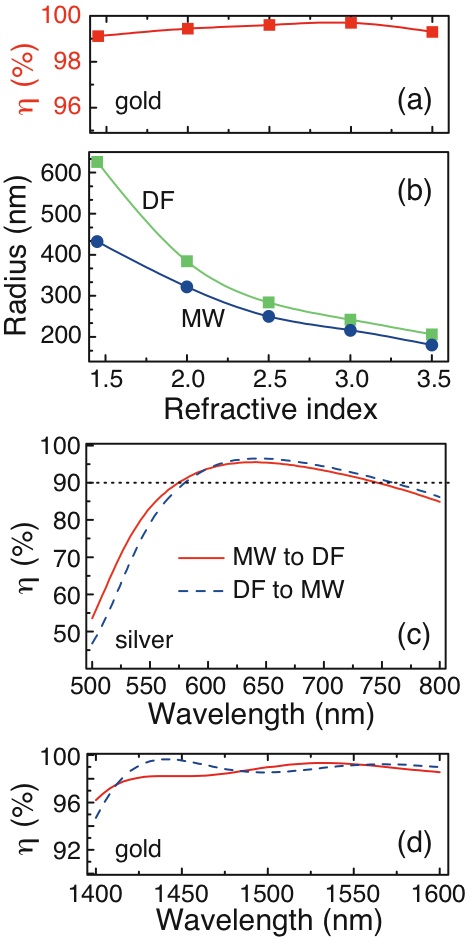}
\caption{(a-b) $\eta$ and optimal radii of a gold MW and DF as a
function of the DF refractive index at $\lambda$=1550~nm.
(c) $\eta$ for a silver MW of radius 164~nm, coupled to a
silica DF of radius 342~nm at $\lambda$=633~nm.
(d) $\eta$ for a gold MW of radius 180~nm
coupled to a silicon DF of radius 208~nm at $\lambda$=1550~nm. \label{fig4}}
\end{figure}

\begin{figure}
\includegraphics[width=16.5cm]{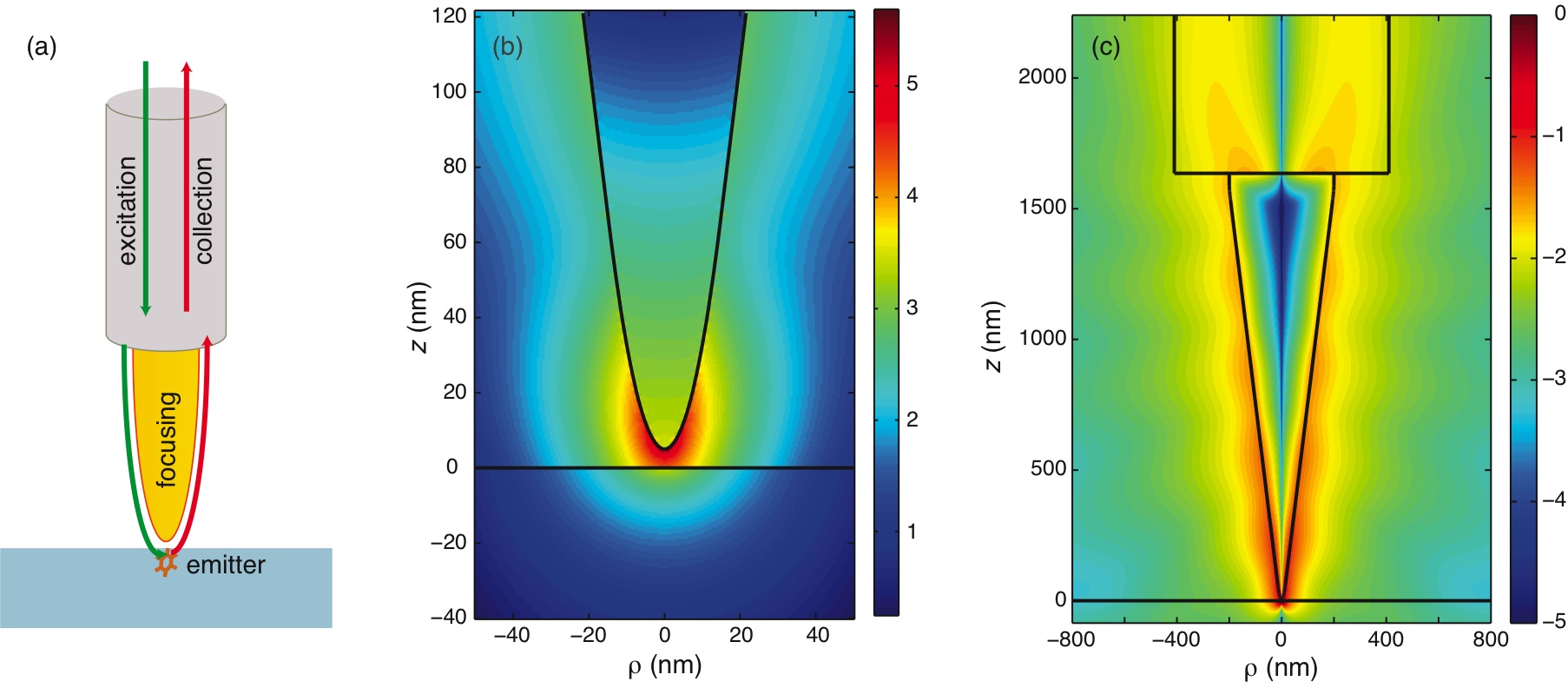}
\caption{(a) Sketch of the device consisting of a
metallic nanocone butt-coupled to a DF. (b)
Normalized distribution (logarithmic scale) of the electric field
intensity near a substrate when photons are launched from the DF into
the MW cone.
(c) Time-averaged magnetic field
radiated by a single emitter embedded in the substrate.
See text for details. \label{fig5}}
\end{figure}

Variation of the DF material can be used to engineer the
coupling device to match the confined SPP modes on thin MWs.
As inferred from \ref{fig2}(a) and (b) the optimal radius
of the MW can be reduced by a high-refractive-index
DF also because the SPP standing wave on the coupling interface
gets compressed by the larger wavenumber.
\ref{fig4}(a) and (b) show the mode conversion
properties for a gold MW at $\lambda$=1550~nm as a function of
the refractive index of the DF. The optimal radii of the MW and DF
decrease and converge to nearly the same value for larger DF
refractive indices, while in each case $\eta$ can be optimized
beyond 99\%.
As a last important property of concern, we examined the
bandwidth of $\eta$ for a set of fixed MW and DF radii.
\ref{fig4}(c) displays $\eta$ for a
silver MW coupled to a silica DF in the visible range, and
\ref{fig4}(d) shows the same for a gold MW coupled to a silicon
DF in the near-infrared regime. These results demonstrate that
bandwidths greater than 150~nm are fully within reach for over 90\%
conversion of SPPs to photons and vice versa.

We now discuss an immediate consequence of efficient and broad-band
SPP-photon conversion between a MW and a DF from both directions.
Over the last decade, several reports have pointed out that SPPs can
be focused tightly to nanoscopic regions at metallic
tips\cite{keilmann99,babadjanyan00,stockman04}. This phenomenon can
indeed revolutionize high-resolution scanning near-field optical
microscopy (SNOM), which has suffered from a very low transmission
through small apertures\cite{novotny95b}. However, so far there has
been no viable approach for feeding optical energy into the SPP mode
of the MW with a high efficiency. The device concept sketched in
\ref{fig5}(a) provides an ideal solution for simultaneous intense
and localized illumination as well as efficient collection.
Furthermore, it is fully compatible with both state-of-the-art
nanofabrication\cite{deangelis08} and scanning implementation of
fluorescence, Raman, or other nonlinear
nanoscopies\cite{sanchez99,ichimura04}.

As a concrete example, we considered illumination at
$\lambda$=730~nm and fluorescence collection at $\lambda=$780~nm
from an emitter with a transition dipole moment oriented along the
tip axis and embedded 5~nm below surface of a sample with a
refractive index of 1.7. The excitation light was launched from a DF
of 410~nm radius, coupled to a gold nanocone with an initial radius
of 200~nm, and an opening angle of 14 degrees, which was terminated
by a paraboloid. The tip was kept at a separation of 5~nm from the
substrate. In \ref{fig5}(b) we plot the electric field
intensity normalized to that of the source in the DF on a
logarithmic scale for a zoom of a small region around the tip.
The intensity is enhanced by more than 5 orders of magnitude
and exhibits strong
localization. \ref{fig5}(c) illustrates the emission process.
We find that the radiated power is funnelled
in the SPP and is then converted into the
TM$_{01}$ mode of the DF with an overall collection efficiency
of about 70\%, which is only limited by propagation loss and
radiation along the nanocone.

We have shown that confined SPPs of a MW can be converted into
guided photons of a dielectric nanofiber with a very high efficiency
and large bandwidth using a simple butt-coupling scheme.
For a given wavelength
between the visible and the near infrared spectral range,
one can always obtain $\eta$
larger than 95\% by an appropriate choice of radius and material of the
MW and DF. We have found that moulding of SPPs at the MW-DF interface
plays a fundamental role in achieving these performances.
Furthermore, we have discussed an important application of butt-coupling
in the context of SNOM and nanofocusing, proposing a scanning probe that
combines broad-band and high-throughput with high-spatial resolution.
Efficient, broad-band and low-loss conversion of
confined SPPs to guided photons
is also key for a range of other applications such as
high-resolution colour imaging\cite{kawata08}, on-chip manipulation
and processing of quantum optical
signals\cite{domokos02,chang06,chang07} for implementation of
quantum networks\cite{cirac97}, and actuation or sensing of
physical, chemical, or biological processes at the molecular
level,\cite{zheng09} paving the way for {\it molecular-scale
plasmonics}.

We thank F. De Angelis and E. Di Fabrizio for fruitful discussions. This
work was supported by ETH Zurich grant TH-49/06-1.

\end{document}